\documentclass[twocolumn,superscriptaddress, prx]{revtex4-2}

\usepackage{graphicx}
\usepackage{dcolumn}
\usepackage{bm}
\usepackage{amsmath}
\usepackage{textgreek}
\usepackage{makecell,tabularx}
\setcellgapes{8pt}

\begin{document}

\title{Multi-pass guided atomic Sagnac interferometer for high-performance rotation sensing}


\author {Samuel Moukouri} 
\affiliation{ Department of Physics, Ben-Gurion University of the Negev, Be'er Sheva 84105 Israel} 
\author {Yonathan Japha}
\affiliation{ Department of Physics, Ben-Gurion University of the Negev, Be'er Sheva 84105 Israel}   
\author {Mark Keil}
\affiliation{ Department of Physics, Ben-Gurion University of the Negev, Be'er Sheva 84105 Israel} 
\author {Tal David} 
\affiliation{Israel National Quantum Initiative, 43 Jabotinsky, Jerusalem}
\author {David Groswasser}
\affiliation{ Department of Physics, Ben-Gurion University of the Negev, Be'er Sheva 84105 Israel}  
\author {Menachem Givon} 
\affiliation{ Department of Physics, Ben-Gurion University of the Negev, Be'er Sheva 84105 Israel} 
\author {Ron Folman} 
\affiliation{ Department of Physics, Ben-Gurion University of the Negev, Be'er Sheva 84105 Israel}

\begin{abstract}

Matter-wave interferometry with atoms propagating in a guiding potential  is expected to provide compact, scalable and precise inertial sensing.  However, a rotation sensing device  based on the Sagnac effect with atoms guided in a ring has not yet been implemented despite continuous efforts during the last two decades. Here we discuss some intrinsic effects that limit the coherence in such a device and propose a scheme that overcomes these limitations and enables a multi-pass guiding Sagnac interferometer with a Bose-Einstein condensate (BEC) on a chip in a ring potential. We analyze crucial dephasing effects: potential roughness, phase diffusion due to atom-atom interactions and number uncertainty, and phase fluctuations. Owing to the recent progress in achieving high momentum beam splitting, creating smooth guides, and manipulating the matter-wavepacket propagation, guided interferometry can be implemented within the coherence time allowed by phase diffusion. Despite the lower particle flux in a guided Sagnac ring and the miniaturization of the interferometer,  the estimated sensitivity, for reasonable and practical realizations of an atom chip-based gyroscope, is comparable to that of  free-space interferometers, reaching $\text{45}\,\text{nrad}\,\text{s}^{\text{-1}}\,\text{Hz}^{\text{-1/2}}$. A significant improvement over state-of-the-art free-space gyrocope sensitivities can be envisioned by using thermal atoms instead of a BEC, whereby the interferometer can be operated in a continuous fashion with the coherence limited by the scattering rate of the atoms with the background gas. Taking into account the sensitivity times length of the interferometer as the figure of merit which takes into account compactness, our configuration is expected to deliver a potential improvement of 2-4 orders of magnitude over state-of-the-art free-space gyroscopes for a BEC, and 4-6 orders of magnitude for thermal atoms.

\end{abstract}

\maketitle

\section{ Introduction}

Ultra-sensitive rotation sensors have become a primary goal for fundamental physics as well as practical devices such as inertial navigation systems. Achieving high sensitivity is a means for refining fundamental constants and to confirm or falsify competing theories in quantum mechanics, general relativity, and geophysics \cite{stedman,schreiber}. For everyday applications, inertial navigation systems are especially important for GPS-denied environments such as deep oceans, urban areas, or in situations where the GPS is jammed, spoofed, or hacked. Improvements of mechanical or optical sensors have been slow during the last decade. Constructing an inertial navigation system based on atomic interferometers is currently the goal of different groups around the world \cite{barrett,gustavson,kovachy,gauguet,dutta,berg,wu,tonyushkin,schubert,hopkins,baker, jiang1,reeves,jiang2,horne,moan,gupta,qi,lesanovski,sherlock,navez,griffin, vangeleyn,pritchard,heathcote, henderson,turpin,bell,morizot,stevenson}.

Owing to the large ratio of atomic mass to the photon effective mass, matter-wave precision sensors are appealing when compared with those based on light. While this ratio is theoretically predicted to be as large as ten orders of magnitude, in realistic scenarios the gap between the achievable performance is not as large. In the last two decades, atomic gyroscopes in which cold atoms move in free space were built. They are already competitive with the best optical gyroscopes for long term stability and accuracy \cite{barrett}. Despite these successes, there still exists significant unexploited potential in terms of sensitivity and miniaturization. Some important factors hamper progress: atomic fluxes are lower than those of light by more than ten orders of magnitude, resulting in a reduction of more than five orders of magnitude in sensitivity. In addition, the area of the sensors remains very small in atomic interferometers. Furthermore, free-space cold atom interferometers are currently quite large, heavy, consume significant power, are expensive, and are  technologically complex \cite{kovachy}. However, a new multi-loop concept which can significantly improve the performance of free-space interferometers was recently proposed \cite{schubert}. 

Current efforts are devoted to the development of guided gyroscopes. In these new gyroscopes, a ring-shaped potential is generated optically or magnetically. The gyroscope sensitivity can be significantly improved without increasing the interferometer size because the atoms can, in principle, perform several revolutions in the loop, similar to the way light travels through typical fiber-optic gyros (FOGs) or ring-laser gyros (RLGs). Several theoretical and experimental configurations for guided atom gyroscopes have been the subject of investigations in the last two decades. These configurations include magnetically guided rings with static fields \cite{hopkins}, DC currents \cite{baker, jiang1}, DC currents with a time orbiting potential \cite{reeves,jiang2,horne}, time orbiting ring traps \cite{gupta}, coils generated traps \cite{qi}, time-averaged adiabatic potentials (TAAP) \cite{lesanovski,sherlock,navez}, inductively coupled rings \cite{griffin, vangeleyn,pritchard}, optical dipole potentials \cite{heathcote, henderson,turpin,bell}, and magnetic potentials coupled to light   \cite{morizot}. 

A different interferometer concept was proposed in Ref.\,\cite{stevenson}. In this interferometer, the atoms do not move in a guide, but in contrast remain trapped, and the traps are  moved using spin-dependent potentials. The advantage of this type of interferometer is that the wavepackets do not spread as in a guide and hence there is {\it a priori} no loss of atoms during the operations.  In addition, the plane of the ring may be placed parallel to gravity (for rotation sensing on other axes) without the hindering effect of gravity slowing down the atoms. However, the acceleration of the traps by the time-dependent potentials is a new source of noise which is absent in guided interferometers. 

Magnetic or dipole rings have been realized experimentally by several groups. Others have constructed moving straight guides in which the effective motion of the atoms in the lab frame is a closed loop  allowing rotation measurement \cite{wu, tonyushkin,garcia,burke}. The loading of ring guides and  the performance of several revolutions have been reported \cite{gupta,pandey}. A coherence time of up to five seconds during the condensate motion in the ring trap was observed in Ref.\,\cite{pandey}. To our knowledge, a full interferometric cycle in a ring:  loading the atomic cloud, and splitting and recombining the wavepackets, has not been realized. We note that a full interferometric cycle was performed in Refs.\, \cite{wu,tonyushkin} with a moving straight guide. However,  the effective Sagnac area and the sensitivity to rotation ($\text{0.12}\,\text{mm}^{\text{2}}$ and $\text{4}\,\text{mrad}\,\text{s}^{\text{-1}}\,\text{Hz}^{\text{-1/2}}$) were far from the performance of the best free-space interferometers in which Sagnac areas as large as $\text{24}\,\text{mm}^\text{2}$ and  sensitivities below $\alt \text{10}^{\text{-7}}\, \text{rad}\, \text{s}^{\text{-1}}\, \text{Hz}^{\text{-1/2}}$ were obtained \cite{barrett}.  To the best of our knowledge, the state-of-the-art in moving linear guides, eventually enclosing an area, is currently set by the Boshier group in Los Alamos, where 200 round trips were reported in an enclosed area of 2\,mm$^2$, with a coherence time of 1\,s \cite{krzyzanowska}. This work already constitutes a big step forward in the world effort.

The bottleneck in constructing a guided matter-wave gyroscope stems from several hindering effects which dramatically affect the performance of the interferometer. The first effect is guide roughness. Guides from current-carrying wires or light potentials suffer from roughness due to wire imperfection or laser noise. This was a serious limitation for atom manipulation since roughness could lead to the fragmentation of the atom cloud and reflections. Improvements in fabrication and implementation of smoothing methods based on time averaging have yielded very smooth guides \cite{folman,trebbia,bouchoule}. In a recent TAAP study, a guide roughness of less $\text{200}\,\text{pK}$ was achieved \cite{pandey}. The atoms were moved coherently at supersonic speeds up to the BEC coherence time of about 5\,s.  

The second effect is phase diffusion induced by atom-atom interactions. Free-space interferometry allows  a waiting time long enough $(\sim \text{10}\, \text{ms})$ after the atoms are released from the trap to ensure that interactions do not appreciably affect the interferometer signal. However, in a guide the atoms remain trapped in the directions transverse to their motion. Therefore, the interaction effects in guided interferometry with a BEC have a substantial impact on the output signal. They lead to phase diffusion and loss of coherence between the two parts of an initial wavepacket. 

Dynamical effects are another source of decoherence, and they originate from field fluctuations. For instance, they may paradoxically arise from the driving fields introduced for roughness reduction. The usual criterion, \textOmega$_{\text{driving}} \gg$ \textomega$_{\text{trap}}$ for the validity of the adiabatic limit where the atoms move in an effective static potential is not rigorous \cite{sandberg}. It is more accurate to use the fidelity between the time-averaged state and the ground state of the driving Hamiltonian. The fidelity was found to decay quadratically for short times and exponentially for longer times \cite{sandberg}. This means that dynamical effects are important even when the driving frequency is large with respect to the trap frequency. 

The challenge of building a working guided interferometer lies in the minimization of these impeding effects. The first step is the choice of a guide. It is  preferable to choose guides resulting from static fields for which the main source of randomness is static roughness. Time-dependent fields are  effective in producing very smooth guides, but because they are a source of uncontrollable dynamics, their application should be minimized.  The guide parameters must then be adjusted in order to minimize  roughness and phase diffusion. Finally, owing to recent progress in high-momentum beam splitters \cite{abend}, guided atom interferometers with high sensitivity can be realized. With such high-momentum beam splitters, the effect of gravity on propagation within a ring in a plane parallel to the direction of gravity is also minimized.

In this study, we perform simulations of two ring guides created by static fields: a static magnetic field Sagnac (M-Guide) and an electro-magneto static Sagnac (E-guide). The M-guide is a loop magnetic guide generated by an Archimedean spiral circuit \cite{jiang1,jiang2}. The E-guide is a loop resulting from an electrostatic field of a ring capacitor and a magnetic mirror \cite{hopkins}. 

Our analysis was performed with a BEC in the guide, hence the sensitivity is limited by the available fluxes and phase diffusion effects which limit the interferometer time less than 1\,s. The projected sensitivity is $\text{45}\,\text{nrad}\,\text{s}^{\text{-1}}\,\text{Hz}^{\text{-1/2}}$. Considering the figure of merit of sensitivity times the length of the interferometer, our design shows a potential improvement of 2-4 orders of magnitude over present state-of-the-art free-space interferometers. 

Our ultimate goal, whose proof-of-principle was demonstrated in \cite{yoni}, is to operate the interferometer in a continuous way with a thermal beam formed by a 2D MOT, utilizing tunneling barriers as the beam splitters. In this scenario we may fully benefit from the long vacuum-limited lifetime of about 10\,s for cold atoms in the ring trap. This interferometer configuration, with an atomic flux of
 $\text{10}^{\text{7}}\,\text{s}^{\text{-1}}$, is predicted to yield a sensitivity of $\text{1}\, \text{prad}\,\text{s}^{\text{-1}}\,\text{Hz}^{\text{-1/2}}$ for $10^3$ turns. Again, considering the same figure of merit, the improvement over the state-of-the-art for a thermal beam is potentially up to 4-6 orders of magnitude.

Responding to the challenges of realizing a multi-pass Sagnac ring guide, we discuss design characteristics for the two static potential guides in Sec.~II, technical imperfections of these guides in Sec.\,III, and intrinsic dephasing effects in Sec.\,IV. We show how a BEC released in free space is notably different from release in a guide. Our results from Secs.\,III and IV together therefore allow us to present optimized guide configurations in Sec.\,V, followed by simulations of our expected coherence for realistic traps and wavepacket parameters in Sec.\,VI. Also in Sec.\,VI, we show the projected sensitivities for the M-guide and the E-guide and compare them to state-of-the art sensitivities for free-space interferometers. In Sec.\,\ref{revivals}, we discuss the possibility of avoiding phase diffusion effects and improving the sensitivity owing to quantum revivals of wavepackets.

{\parindent=0pt
\begin{figure*} 
\begin{centering}
  \includegraphics[width=0.67\textwidth]{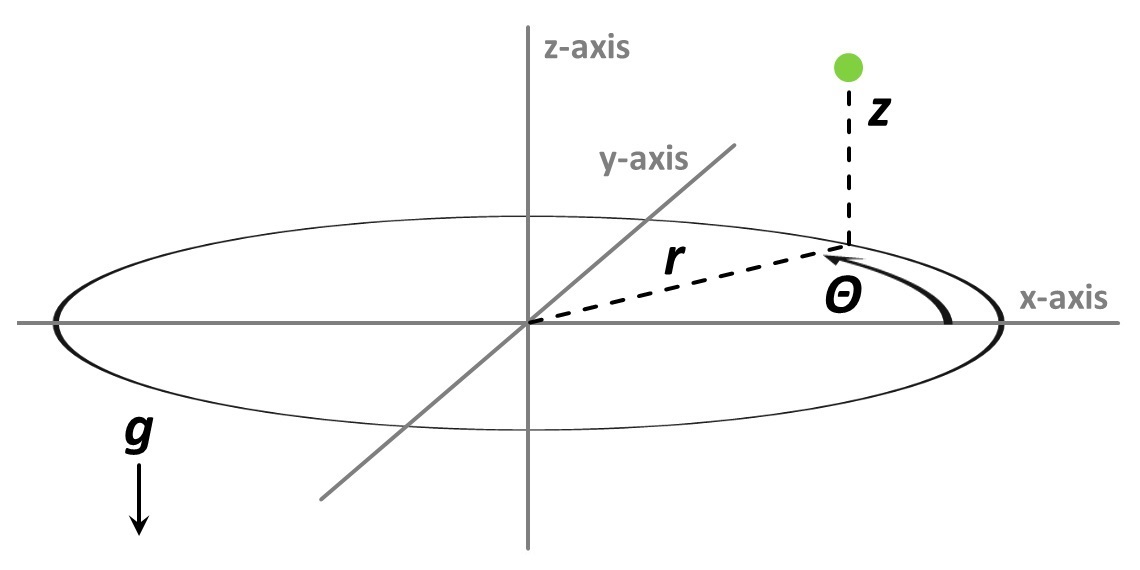}
	\parbox{0.9\textwidth}{\caption{\label{coord-fig} \baselineskip=0.67\baselineskip Co-ordinate system for the ring guides, with gravity pointing in the direction of the arrow labeled~$g$. Radial and axial frequencies are in the~$r$ and~$z$ directions, respectively. The green dot is an illustration of a point in the reference system with its co-ordinates $r,\,\theta,\,z$.}}
\end{centering}
\end{figure*}
}

\section{Guides: M-guide and E-guide}
\label{guides}

In this section, we present the two static guides which will be the subject of our study. The M-guide is a loop magnetic guide generated by an Archimedean spiral circuit \cite{jiang1,jiang2}. The E-guide is a loop resulting from an electrostatic field of a ring capacitor and a magnetic mirror \cite{hopkins}. Fig.\,\ref{coord-fig} shows the co-ordinates used for the ring guides, with gravity pointing in the direction of the arrow. Radial and axial frequencies are in the~$r$ and~$z$ directions, respectively. Perfectly smooth potentials would be characterized by an azimuthal frequency of zero in the $\theta$ co-ordinate.

{\parindent=0pt
\begin{figure*} 
\begin{centering}
   \includegraphics[width=0.75\textwidth]{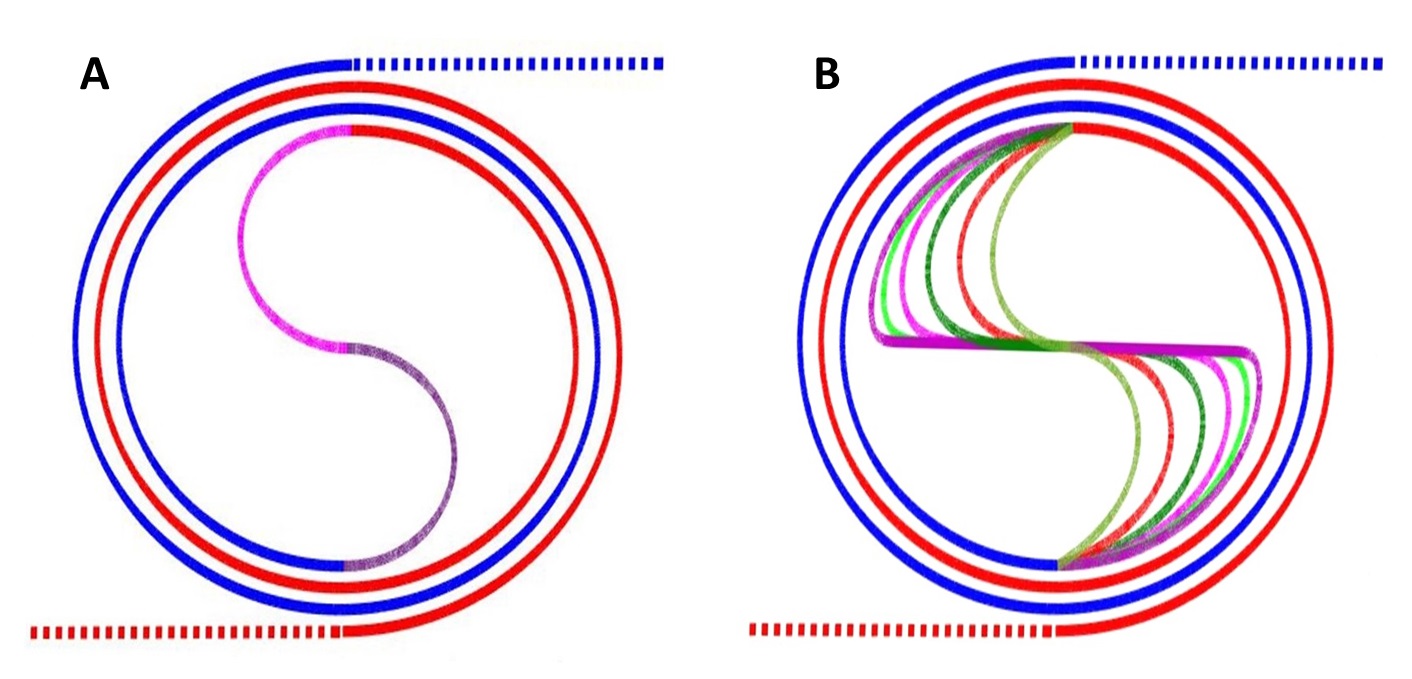} 
	\parbox{0.9\textwidth}{\caption{\label{archi1-fig} \baselineskip=0.67\baselineskip Archimedean spiral with semi-circle and spiral connections: (A)~Archimedean circuit showing two interleaved spirals in blue and red continuous lines. The connection in the middle is made by two semi-circles in purple and magenta. The resulting counter-propagating currents create a 3-wire quadrupole trap around the ring. The external leads are shown by dotted lines. (B)~Modified Archimedean circuit, joined in the middle by spirals whose index~$n$ in the polar equation~$r=\alpha\theta^{1/n}$ is (starting from the green curve in the center) {$n=2, 4, 8, 16, 32, 64$}. Even values of~$n$ generate two branches (e.g.,~$n=2$ is the Fermat spiral), while~$n=1$ is the single-branch Archimedean spiral.}}  
\end{centering}
\end{figure*}
}

\subsection{M-Guide: An Archimedean circuit generated magnetic guide}

The Archimedean spiral DC circuit used to generate a magnetic ring guide was proposed in Ref.\,\cite{jiang1}. The circuit generating the M-guide is made of two interleaved Archimedean spirals which are connected by two semi-circles as shown in Fig.\,\ref{archi1-fig}(A). In Ref.\,\cite{jiang1}, the current was set to $I$=2\,A, yielding a deep trap of about 400\,\textmu K with radial and axial frequencies of approximately 20\,kHz. The ring guide is located at a height of 102\,\textmu m from the circuit.  Two important limitations of this guide, shown in Fig.\,\ref{spirals-fig}(A) for $I$=1\,A, are the large potential bumps (approximately 75\,\textmu K) induced by the input and output leads and two zero potential points in the valleys between these bumps at $\theta=90^\circ$ and $\theta=270^\circ$. 

The potential zeroes must be eliminated in order to prevent atom loss due to Majorana spin flips, while the potential bumps must obviously be made significantly smaller than the wavepacket's kinetic energy. The former can be eliminated by using different curves as junctions between the two spirals. We find that the minima shift to non-zero values by choosing segments of spirals instead of the semi-circles used in the original circuit. We use the general class of spirals whose equation in polar co-ordinates is given by $r=\alpha\theta^{1/n}$, where $\alpha$ is a real number and $n$ is an integer; note that the Archimedean spiral itself corresponds to $n=1$, while $n=2$ is known as the Fermat spiral. We show in Fig.\,\ref{archi1-fig}(B) the two interleaved Archimedean spirals with central spiral connections for different values of $n$. We find that as $n$ increases, the zeros of the potential are progressively shifted to larger values. The highest non-zero minimum is reached near $n=64$ which is shown in Fig.\,\ref{spirals-fig}(B). For $n\,>\,64$, the potential minimum decreases and seems to return to zero as $n$ goes to infinity.

{\parindent=0pt
\begin{figure*} 
\begin{centering}
   \includegraphics[width=0.9\textwidth]{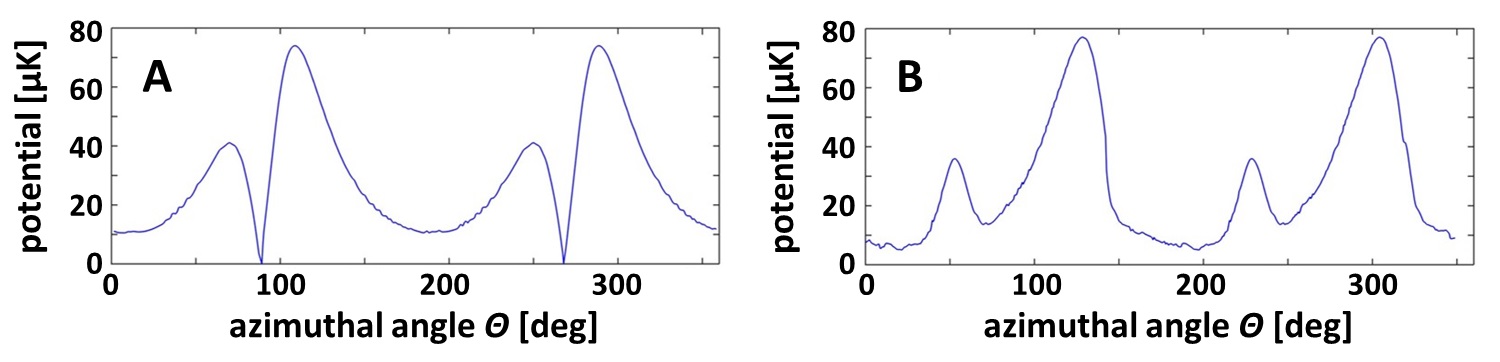} 
	\parbox{0.9\textwidth}{\caption{\label{spirals-fig} \baselineskip=0.67\baselineskip Archimedean spiral M-guide potential: (A)~The potential around the ring displays large bumps and zeros. The zeros are located near~$\theta=90^\circ$ and~$\theta=270^\circ$ where the magnetic field changes sign (only the absolute value is plotted here). (B)~The potential zeros are lifted at~$\theta=90^\circ$ and~$\theta=270^\circ$ by replacing the central connection with the~$n=64$ spiral of Fig.\,\ref{archi1-fig}(B).}}
\end{centering}
\end{figure*}
}

{\parindent=0pt
\begin{figure*} 
\begin{centering}
   \includegraphics[width=0.9\textwidth]{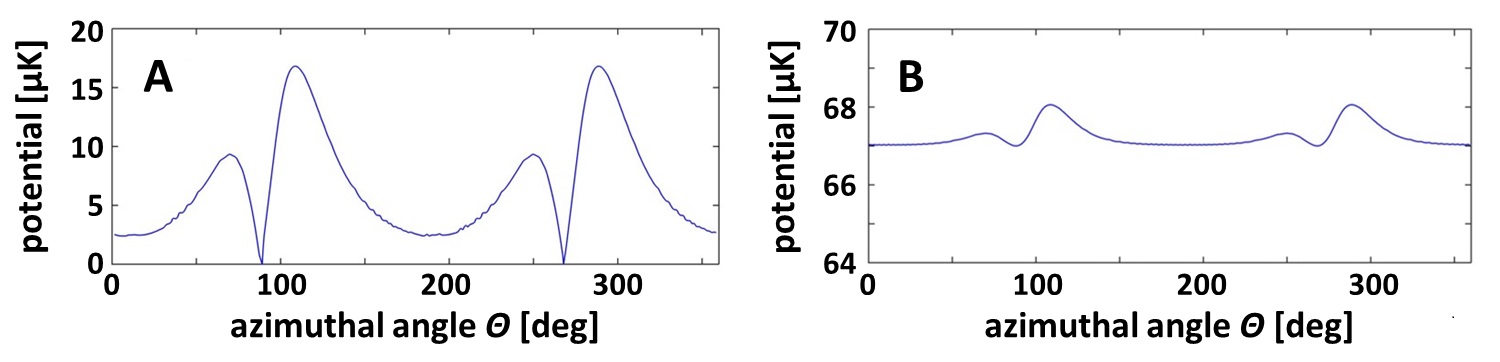} 
   \parbox{0.9\textwidth}{\caption{\label{archiTOP-fig} \baselineskip=0.67\baselineskip Smoothing the Archimedean spiral M-guide potential by adding a TOP field: guide potentials for (A)~B$_{\rm TOP}=0$ and (B)~B$_{\rm TOP}=\rm1\,G$. Note that the spiral has been rescaled by a factor of~4.4 (see text).}}
\end{centering}
\end{figure*}
} 

A more effective method for eliminating the zeros and reducing the bumps was proposed in Ref.\,\cite{jiang2}. It uses coils to create a time-orbiting potential (TOP) \cite{petrich}. The TOP field shifts the trap minimum away from the zeros and yields very smooth potentials. When the distance from the trap is 102 \textmu m and the current is $I$=1\,A, the bump height at zero TOP field is 75\,\textmu K (1.12\,G). If the TOP field is equal to 5\,G,  the bump is reduced to 4.2\,\textmu K. Given that the energy of a $^{87}$Rb atom with a velocity of $\hbar k$ is 0.18\,\textmu K ($k=2\pi$/780\,nm corresponds to the $\text{D}_2$ line), the atoms must have a velocity of at least 5\,$\hbar k$ to pass the bump. This is still challenging for cold-atom manipulations. Smoother guides can be obtained by modifying the distance between the spirals. If we rescale  the spiral by a factor of 4.4, the height of the guide shifts to 445\,\textmu m from the chip. As shown in Fig.\,\ref{archiTOP-fig}(A), the bump magnitude drops to 17\,\textmu K. At this height, adding a TOP field of 1\,G reduces the bump magnitude to 1.02\,\textmu K as shown in Fig.\,\ref{archiTOP-fig}(B). An atom having a velocity of 3\,$\hbar k$ will pass over such potential bumps.

Since the motion of the zeros is shifted from the trap bottom to the so-called circle of death \cite{petrich}, the trap depth significantly decreases when the TOP field is applied. In the case above, when $I$=1\,A, it goes from 200\,\textmu K  to only 10\,\textmu K when the TOP field magnitude is 5\,G. Higher TOP fields can be applied to the circuits with spiral connections will yield deeper traps because in that case there is no circle of death.

\subsection{E-Guide: A capacitor-magnetic mirror generated guide}

{\parindent=0pt
\begin{figure*} 
\begin{centering}
   \includegraphics[width=0.9\textwidth]{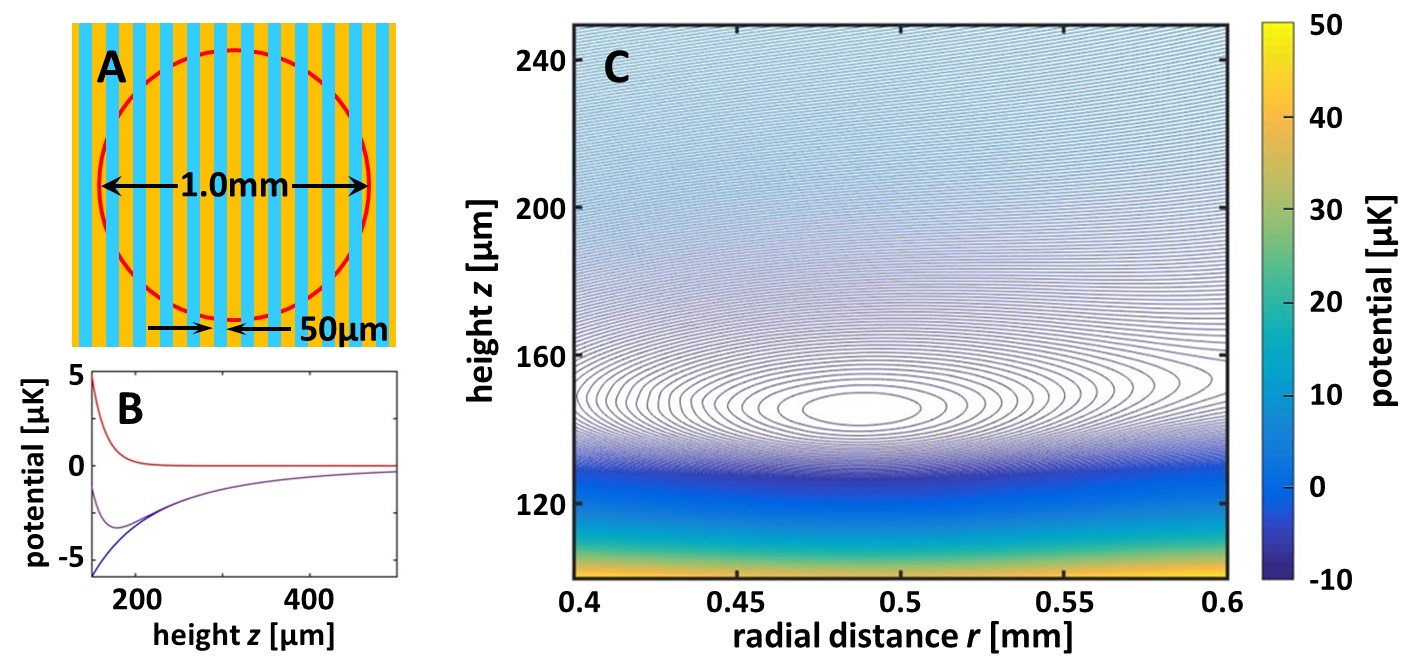} 
	\parbox{0.9\textwidth}{\caption{\label{mirror-fig} \baselineskip=0.67\baselineskip (A)~Sketch of the central portion of the E-guide atom chip. The red ring is a~$\rm50\,\mu m$-gap in a charged~$\rm2\,\mu m$-thick gold surface, making it a capacitor and providing the attractive part of the potential. The blue stripes are thin-film magnetized layers, with a period of~$\rm100\,\mu m$ as shown, forming the repulsive potential (a smooth magnetic mirror, see Ref.\,\cite{roach}). The gold conductor and magnetic stripe structures of the complete atom chip are much larger than depicted here, so the potential is expected to be axially symmetric. (B)~The attractive electric potential (blue curve) generated by a capacitor with voltage of~$\rm100\,V$, the repulsive potential (red curve) generated by the magnetic mirror with a surface magnetization of~$\rm1000\,G$, and the total potential (purple curve). (C)~Field lines of the E-guide potential in the presence of gravity and the centrifugal force acting on a particle with a velocity of 6\,$\hbar k$.}}  
\end{centering}
\end{figure*}
}

The second guide of our study is created by combining the attractive potential of an electrostatic disc capacitor with the repulsive
potential of a static magnetic mirror \cite{roach}. We will refer to this guide as the E-guide, first proposed in Ref.\,\cite{hopkins}. The E-guide has the crucial advantage of requiring no currents, a known source of noise, as well as requiring only small amounts of conducting metal, reducing the level of Johnson noise. A sketch of the E-guide chip is displayed in Fig.\,\ref{mirror-fig}. The attractive electric potential is generated by applying a potential difference across a 50\,\textmu m-wide circular gap fabricated in a gold layer, thereby creating a strong electric field at the gap edges. This field decays exponentially with the distance from the gap. The magnetic mirror is a grooved structure of a magnetic material with a periodicity of 100\,\textmu m that generates a magnetic field whose modulus decays exponentially away from the mirror surface. These two layers are separated by a thin layer of a non-magnetic insulator. This configuration requires the insertion of a via which connects the gold layer to an electric potential source, which we place along the symmetry axis of the capacitor. The electrostatic attraction, combined with the magnetic repulsion for weak-field seeking atoms, creates a trap whose distance from the chip surface can be controlled by the capacitor voltage and the magnetization of the grooved structure. The trap has the same geometry as the circular gap in the gold layer. In Fig.\,\ref{mirror-fig} we show the field lines of the E-guide potential created by a capacitor of 0.5\,mm radius and a voltage difference of 100\,V, together with a magnetic field at the mirror's surface of 1000\,G.

\section{Effects of  roughness}
\label{roughness}

Guides from current-carrying wires suffer from roughness due to wire imperfections. This was a serious limitation on atom manipulation with such guides. Experiments show that a rough potential can lead to the fragmentation of the wavepackect \cite{leanhardt1,leanhardt2,wu-zhou}. A solution to this problem was proposed in Ref.\,\cite{trebbia}. This solution consisted in adding a modulation to the trap potential. For a proper modulation frequency, the atoms will not experience roughness but instead they will see an effective smooth potential. However, as discussed in Ref.\,\cite{bouchoule}, the frequency of the modulation should be chosen carefully. If the modulation is too slow, this will give rise to heating. And if the modulation is too fast, the atoms will not be able to follow the instantaneous field; they will experience Majorana spin flips and be lost from the trap.

Given that guide roughness also leads to decoherence, we will examine their relation and determine the minimum kinetic energy needed to retain a particular level of coherence for a given time. 

{\parindent=0pt
\begin{figure*} 
\begin{centering}
   \includegraphics[width=0.9\textwidth]{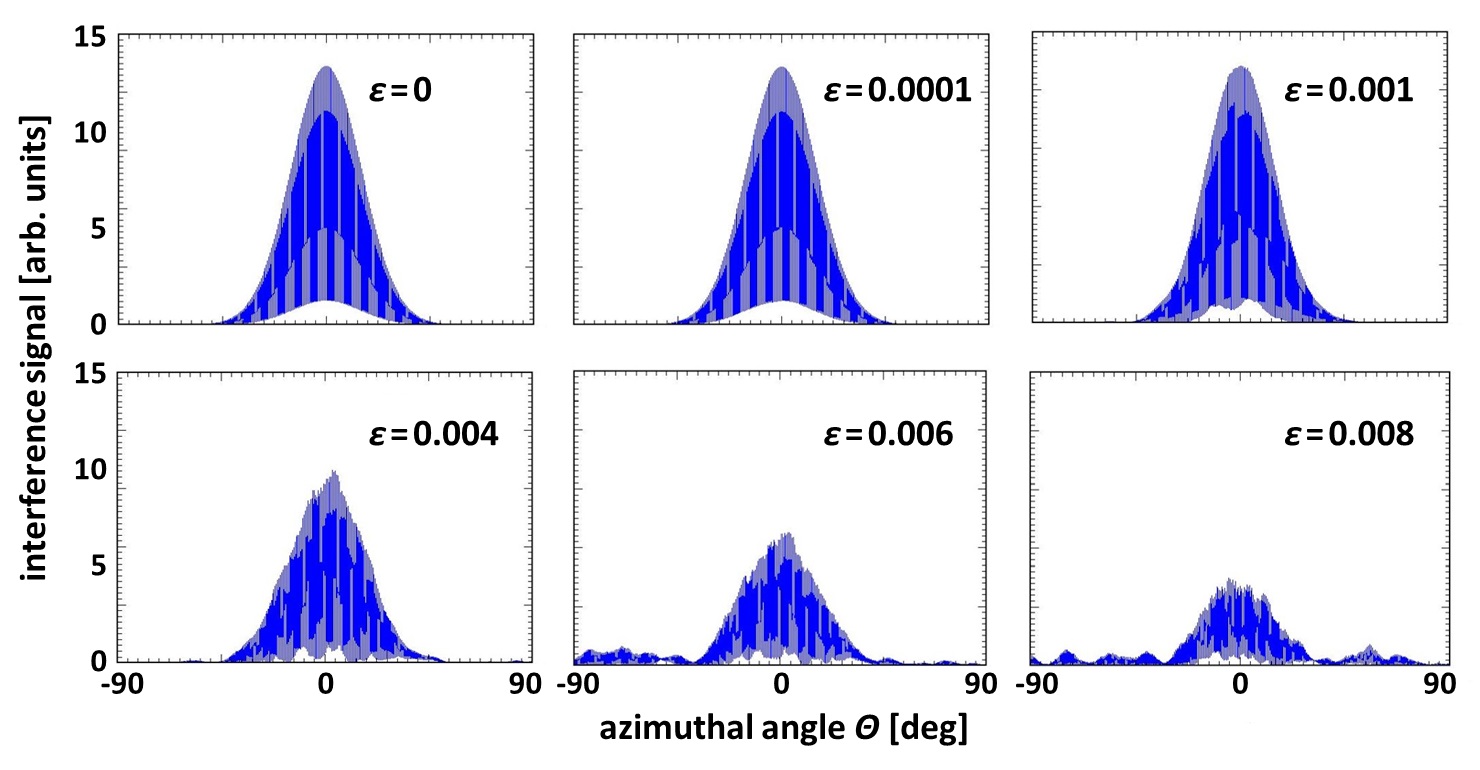} 
	\parbox{0.9\textwidth}{\caption{\label{intrough-fig} \baselineskip=0.67\baselineskip Interference patterns in presence of roughness: $\epsilon=V_R/E_k$ is the ratio between the amplitude of potential variations induced by roughness and the kinetic energy of the atoms. It can be seen that when~$\epsilon<0.001$, the interference pattern is nearly identical to that of~$\epsilon=0$.
When $\epsilon > 0.01$ the contrast becomes small and the interference signal will be difficult to observe.}}
\end{centering}
\end{figure*}
}

We performed numerical simulations of Gaussian wavepackets moving in a ring potential perturbed by roughness modeled by static $1/f$ noise. An estimate of the magnitude of the disorder potential $V_R$, created by this roughness, was calculated in Ref.\,\cite{wang} in the limit where the distance $z_0$ is larger than the width of the wire: 

\begin{equation}
V_R \sim \mu_B \frac{2I}{cz_0} \frac {\delta u \xi^{1/2}}{z_0^{3/2}},
\end{equation}

\noindent where $\mu_B$ is the Bohr magneton, $\delta u$ is the roughness amplitude of the wire, $\xi$ is the noise correlation length at the trap position, and $z_0$ is the distance between the trap and the wire. $V_R$ decays as $z_0^{-5/2}$ which is in agreement with the experimentally observed $V_R \propto z_0^{-2.2}$ \cite{kraft}. We set $2I/cz_0 \sim B_{\text{BOT}}$ to be the field at the trap bottom, with $\xi$=445\,\textmu m and $z_0$=445\,\textmu m. We choose 0.01\,\textmu m $\alt \delta u \alt $1\,\textmu m, typical values of roughness amplitudes obtained by current fabrication techniques \cite{folman,hinds,lucas}. We note here that another source of potential instability would be current fluctuations which, however, would affect only the M-guide since the E-guide does not require any currents at all.  We also point out that the correlation length parameter $\xi$ is much shorter than the width of the bumps previously discussed in the E-guide, whose effects are therefore exponentially damped, i.e., the potential bumps are sufficiently smooth that they will not cause phase decoherence.

In Fig.\,\ref{intrough-fig}, we show the interference pattern of two Gaussian wavepackets in the presence of roughness. We define $\epsilon=V_R/E_k$ as the ratio between the amplitude of potential variation induced by roughness and the kinetic energy of the atoms. When $\epsilon < 10^{-3}$, corresponding, for example, to $\delta u$=0.1\,\textmu m and $E_k=E(4 \hbar k)$, the interference pattern is nearly identical to that of $\epsilon=0$. When $\epsilon > 10^{-2}$ the contrast becomes small and the interference signal will be difficult to observe.
 
In Fig.\,\ref{kickrough-fig}(A), we show the ratio $\epsilon$ of the amplitude of energy fluctuations induced by roughness over the kinetic energy $E(2\hbar k)$ as a function of roughness amplitude for different heights from the chip. When $z_0$=10\,\textmu m, $\epsilon > 10^{-2}$, the results  displayed in Fig.\,\ref{intrough-fig} show that the guide roughness will lead to small contrast. As the guide position is moved away from the chip, $\epsilon$ decreases and ultimately, near the position of the M-guide at $z_0$=445\,\textmu m, $\epsilon < 5 \times 10^{-3}$ up to $\delta u$=0.25\,\textmu m.

{\parindent=0pt
\begin{figure*} 
\begin{centering}
   \includegraphics[width=0.9\textwidth]{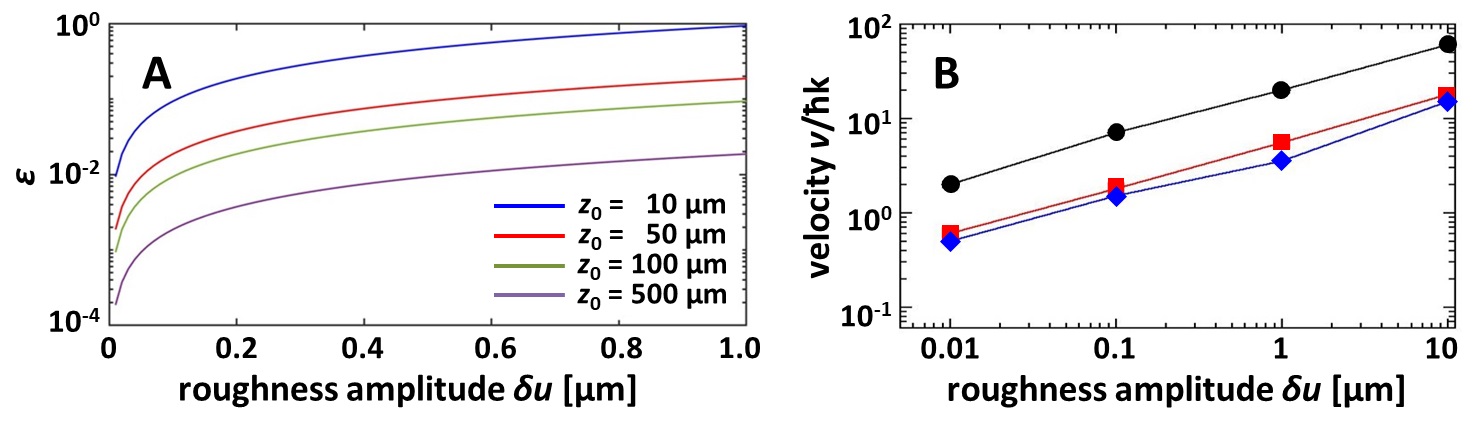} 
	\parbox{0.9\textwidth}{\caption{\label{kickrough-fig} \baselineskip=0.67\baselineskip Roughness magnitude versus kinetic energy: (A)~the ratio~$\epsilon$ of the amplitude of energy fluctuations induced by roughness over the kinetic energy~$E(2\hbar k$) as a function of roughness amplitude for different distances from the chip. At~$z_0$=500\,\textmu m, $\epsilon$ remains below 0.005 up to~$\delta u$=25\,\textmu m. (B)~Atom velocity as a function of roughness amplitude that yields a contrast of 48\%~(diamonds), 75\%~(squares), and 100\%~(circles).}}
\end{centering} 
\end{figure*}
}

In  Fig.\,\ref{kickrough-fig}(B) we show the magnitude of velocity necessary for the atoms to yield a given contrast (48\%, 75\%, 100\%) after one revolution in a ring guide of 1\,mm radius as a function of the roughness amplitude. When the roughness amplitude $\delta u=$0.01\, \textmu m (a value which can be achieved by current fabrication methods), even $v=2\,\hbar k$ is sufficient to yield undiminished interference.

Our analysis shows that when the roughness amplitude is 0.01\, \textmu m, energy fluctuations induced by roughness in the M-guide will be very small and coherence will not be affected. When $\delta u$=0.01\,\textmu m, the amplitude of the energy fluctuation is $V_R=$134\,pK at $z_0$= 500\,\textmu m. Even smaller values of $V_R$ would be obtained by including the effects of the TOP field on roughness, which were not considered in our analysis. Finally, it is known that at a distance $z_0$ from the chip, short-range fluctuations with correlation length $\xi \ll z_0$ are exponentially suppressed \cite{wang,yoni2008}. The suppression of short-range fluctuations which was not included in our calculation will yield even higher coherence. We thus conclude that for the M-guide at about $z_0=$500\,\textmu m, wire roughness will not be an impediment to coherence.

\section{ Phase diffusion in a ring guide}
\label{diffusion}

\subsection{Phase diffusion rate in the time-dependent Thomas-Fermi approximation} 

If a BEC is split into a superposition of two paths, and the two parts are later recombined, the observation of  fringes  resulting from the interference patterns of many single events depends on the length of time during which the two parts travel separately. For short times, the contrast of averaged interference patterns will be high, with the two parts maintaining their phase coherence. However, for long times, phase coherence is lost and the contrast will be small. This diminishing contrast is a manifestation of phase diffusion which is due to atom-atom interactions and to the related number uncertainties \cite{lewenstein,altman,demler,stimming,grond,fallon}, both of which are negligible for free-space interferometers. The theory of phase diffusion, based on the Thomas-Fermi approximation for a BEC in an initial trap that is subsequently split into two identical traps, was derived in Ref.\,\cite{javanainen}. This theory was later extended to ring guides \cite{ilo-okeke}. 

In the Sagnac interferometer, the BEC will be transferred from an initial trap and then allowed to expand in the ring trap where it is later split.
The release from the initial trap and the expansion in the ring guide are usually described by the dynamical Thomas-Fermi approximation of Ref.\,\cite{castin-dum,kagan} which is based on the scaling hypothesis. Here, we will adapt the phase diffusion theory of Ref.\,\cite{javanainen} to the splitting in a guide by using the dynamical Thomas-Fermi approximation. 

Experimentally, a BEC is first prepared in a loading trap which we will assume is isotropic for simplicity. For free-space interferometry, the atoms are then released from this trap, so all the trap frequencies drop rapidly to zero. In contrast, only the azimuthal frequency ($\theta$ co-ordinate in Fig.\,\ref{coord-fig}) drops to zero for interferometry in ring guides. Letting $\lambda_x(t)$, $\lambda_y(t)$, $\lambda_z(t)$ be the scaling parameters for the wavefunction \cite{castin-dum}, the phase diffusion rate $R(t)$ in the dynamical Thomas-Fermi approximation is given by

\begin{equation}
R(t)=\frac{R_0}{\lambda_x(t)\lambda_y(t)\lambda_z(t)},
\label{rate}
\end{equation}

\noindent where $R_0$ is the phase diffusion rate in the static case. Eq.\,\ref{rate} is obtained from the fact that the phase diffusion rate and the chemical potential have the same dependence on the coefficients $\lambda$ \cite{castin-dum, javanainen}. We will now show the difference in behavior of $R(t)$ between the BEC release in free space and in a guide.

{\parindent=0pt
\begin{figure*} 
\begin{centering}
   \includegraphics[width=0.67\textwidth]{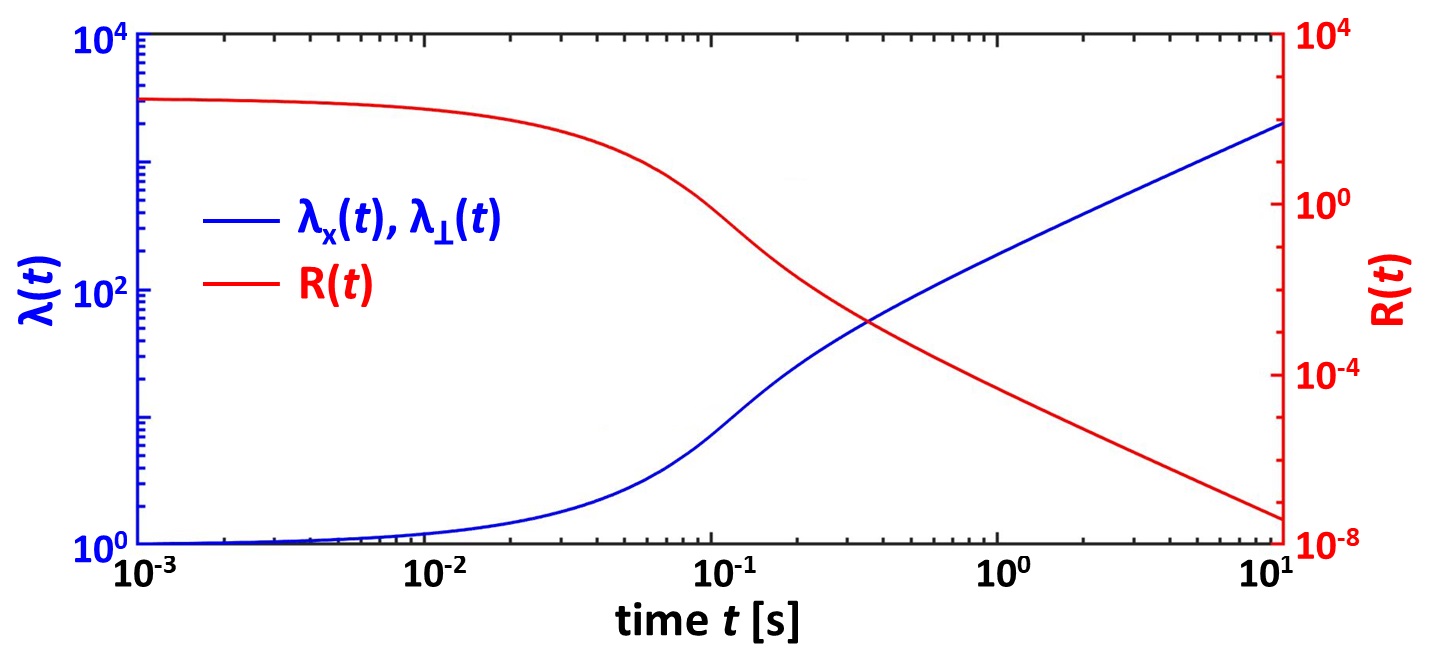} 
	\parbox{0.9\textwidth}{\caption{\label{lambda-iso} \baselineskip=0.67\baselineskip Phase diffusion rate in a free space: numerical solution of the time-dependent Thomas-Fermi equations~\cite{castin-dum} with~$\omega_x(0)=\omega_\perp(0)=\rm2\pi\times1\,kHz$ and final frequencies~$\omega_x=\omega_\perp=0$. Longitudinal ($\lambda_x(t)$) and transverse ($\lambda_\perp(t)=\lambda_x(t)$) wavepacket expansion coefficients are shown (left-hand scale) as a function of time for a trap opening time of~$t_0=\rm10^{-2}\,s$ and~$N=10^5$ atoms. The phase diffusion rate~$R(t)=R_0/\lambda_x(t)^3$ is shown (right-hand scale) after expansion; the initial phase diffusion rate is~$R_0=\rm3.18\times10^2\,s^{-1}$ for the frequencies given above.}}
\end{centering}
\end{figure*}
}

\subsection{BEC release in free space}

 Since we are considering a situation where the initial trap is isotropic and all the directions are opened simultaneously,
$\lambda(t)=\lambda_x(t)=\lambda_y(t)=\lambda_z(t)$. The phase diffusion rate in this case is given by $R(t)=R_0/\lambda(t)^3$. In Fig.\,\ref{lambda-iso}, we show the expansions coefficients $\lambda(t)$ and the phase diffusion rate $R(t)$ after release in free space from an isotropic trap with $\omega= 2\pi \times 1$\,kHz. Fig. \ref{lambda-iso} displays the steep decay of $R(t)$. After a waiting time $t_s$=50\,ms (measured from the beginning of the trap opening), $R(t)$ is already 20 times
smaller than its initial value, and at $t_s$=100\,ms, $R(t)$ is nearly 400 times smaller. $R(t)$ is 10 orders of magnitude smaller than its initial value after 10\,s. The cloud coherence ${\cal C}(t) \approx 1$ during the interferometric cycle (we do not include in this analysis other sources of coherence loss). Hence, after an appropriate waiting time,  phase diffusion does not significantly impact free-space BEC interferometry.

\subsection{BEC release in a ring guide}

The situation is quite different in a ring guide, because the transverse degrees of freedom remain confined after release. This has a significant impact on the coherence. Consider, for example, an initially trapped BEC that is released into a ring guide of radius 1\,mm and subsequently split into two halves that travel in opposite directions in the ring. Wavepacket expansion in the azimuthal co-ordinate will then surely reduce the interaction between the atoms and will lead to lower phase diffusion rates. But will the coherence time be long enough to allow for a working ring interferometer? 

In this case let $\lambda_x(t)$ be the scaling parameter in the direction of the motion and $\lambda_{\perp}=\lambda_y(t)=\lambda_z(t)$. We are supposing the motion to be locally linear because the BEC size at the time of splitting ($\sim$ 10\,\textmu m) is far smaller than the radius of the ring.

{\parindent=0pt
\begin{figure*} 
\begin{centering}
   \includegraphics[width=0.67\textwidth]{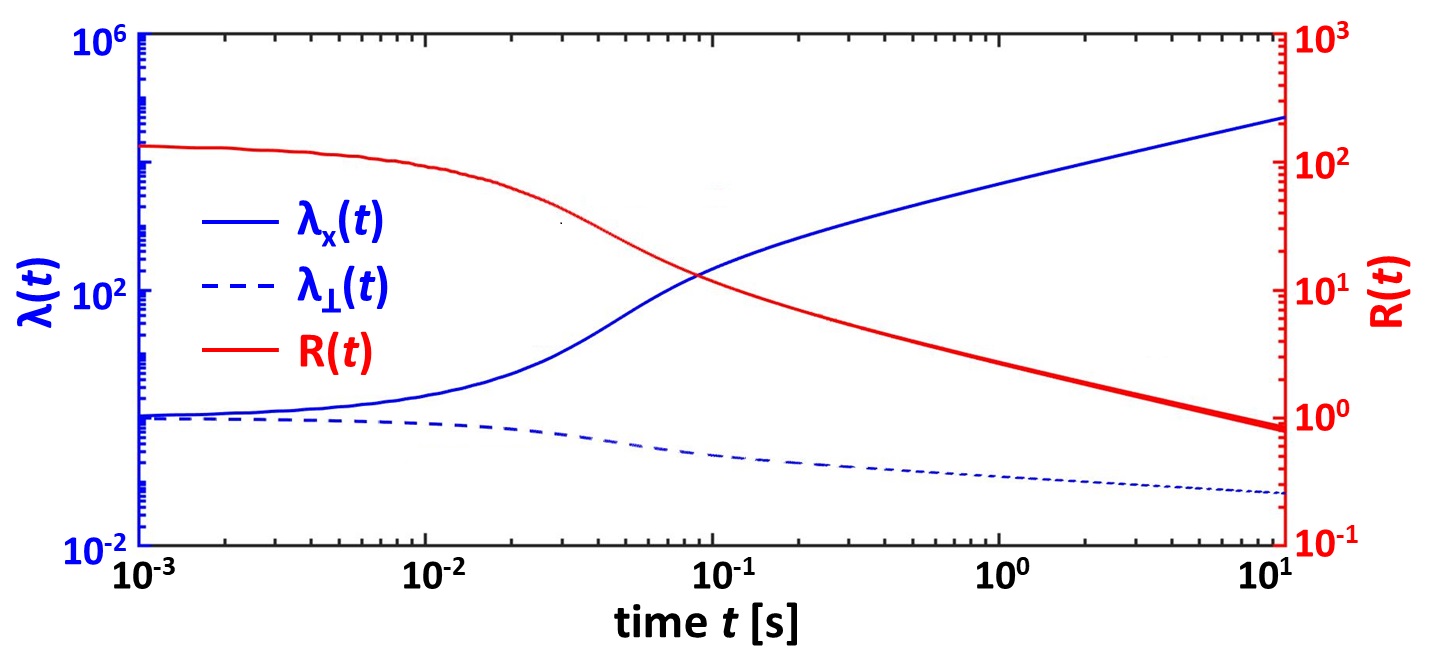} 
	\parbox{0.9\textwidth}{\caption{\label{lambda-aniso} \baselineskip=0.67\baselineskip Phase diffusion rate in a ring guide: numerical solution of the Castin-Dum equations~\cite{castin-dum} with~$\omega_x(0)=\omega_\perp(0)=\rm2\pi\times1\,kHz$ and final frequencies~$\omega_x=0$ and~$\omega_\perp=\rm2\pi\times1\,kHz$. Longitudinal ($\lambda_x(t)$) and transverse ($\lambda_\perp(t)$) wavepacket expansion coefficients are shown (left-hand scale) as a function of time for a trap opening time of~$t_0=\rm10^{-2}\,s$ and~$N=10^5$ atoms. The phase diffusion rate~$R(t)=R_0/\lambda_x(t)\lambda_\perp^2$ is shown (right-hand scale) after expansion; the initial phase diffusion rate is~$R_0=\rm3.18\times10^2\,s^{-1}$ for the frequencies given above.}}
\end{centering}
\end{figure*}
}

In Fig.\,\ref{lambda-aniso}, we display $\lambda_x(t)$ and $\lambda_{\perp}(t)$ after a release in a guide with an average transverse frequency,
$\omega_{\perp}=2\pi \times 1$\,kHz. This frequency was chosen in order to illustrate more dramatically the effects of phase diffusion.
The difference from free-space release is clear. In Fig.\,\ref{lambda-aniso}, for $t_0=10$\,ms, $\lambda_x(t)$ shows rapid growth
from its initial value $\lambda(0)=1$ to more than $1 \times 10^4$ after $t$=10\,s (a duration corresponding to a typical BEC lifetime). At the same time, $\lambda_{\perp}$ decreases from $\lambda_{\perp}(0)=1$ to 0.07. It is this decrease of $\lambda_{\perp}$ that marks the difference with free-space release, for which all the scaling factors~$\lambda(t)$ increase rapidly, thereby effectively suppressing phase diffusion effects. In contrast, the rapid growth of $\lambda_x(t)$ for release into a ring guide is unable to completely overcome the much slower evolution of $\lambda_{\perp}(t)$ (which even decreases) and $R(t)$ does not decay as fast as during a free-space release as seen in Fig.\,\ref{lambda-aniso}. In the ring guide for $t_0=0.01$\,s, $R(t)$ is only 6 and 15 times smaller than its initial value after $t$=50\,ms and $t$=100\,ms, respectively. These factors were 20 and 400 for the corresponding free-space release. It is the slow decrease of $R(t)$ after release in a ring trap that makes guided interferometry so challenging.

{\parindent=0pt
\begin{figure*} 
\begin{centering}
   \includegraphics[width=0.9\textwidth]{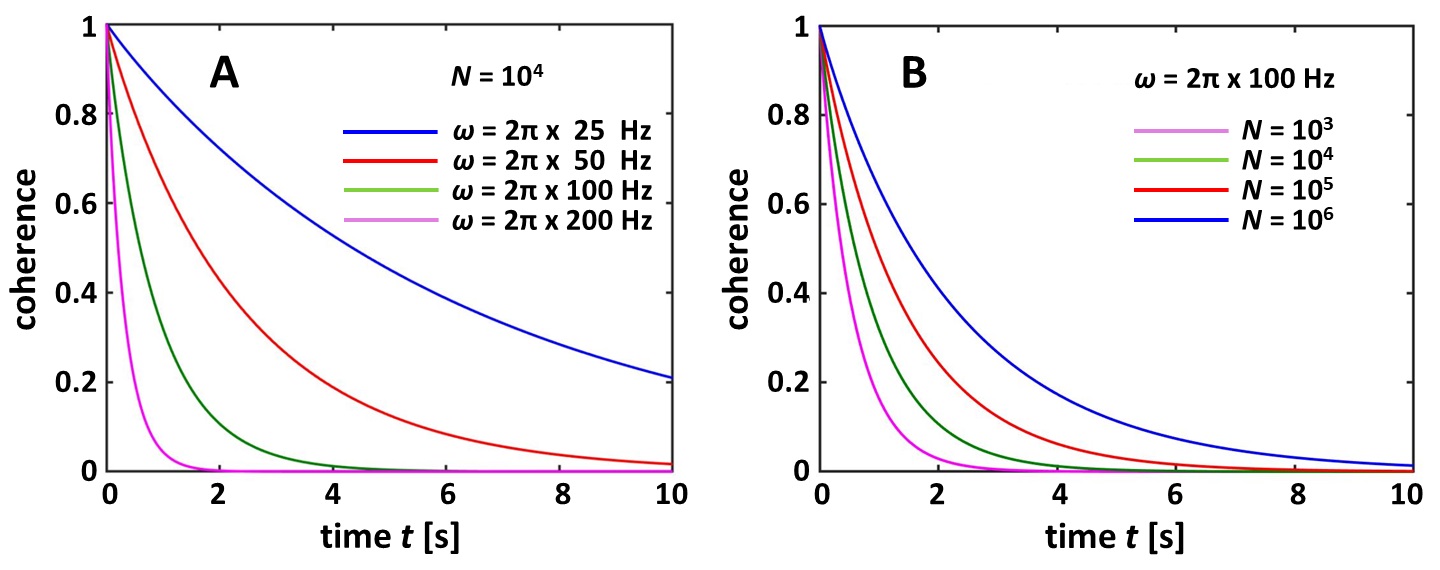} 
	\parbox{0.9\textwidth}{\caption{\label{coherence} \baselineskip=0.67\baselineskip Dependence of the coherence~${\cal C}(t)$ (Eq.\,\ref{coh-eq}) on the trap frequency and the number of atoms. (A)~Dependence on the guide transverse frequency~$\omega$ for~$N=10^4$ atoms; (B)~dependence on the number of atoms~$N$ for~$\omega=\rm2\pi\times100\,Hz$. In all cases the trap opening time is~$t_0=\rm0.001\,s$ and the splitting is performed without any waiting time after the trap opening,~$t_s=\rm0\,s$.}} 
\end{centering}
\end{figure*}
}

 From the phase diffusion rate we can obtain the coherence ${\cal C} (t)$ which is defined as follows,

\begin{equation}
  {\cal C}(t)=\exp \left( -\frac{\Delta \varphi^2(t)}{2} \right),
  \label{coh-eq}
\end{equation}  

\noindent  where $\Delta \varphi(t)$ is given by

\begin{equation}
  \Delta \varphi(t)= R_0 \int_{t_s} ^{t} \frac {du}{\lambda_x(u)\lambda_{\perp}^2(u)}
  \label{phase_2}
\end{equation}

\noindent and $t_s$ is the splitting time.

In Fig.\,\ref{coherence}(A), we show coherence ${\cal C}(t)$ as a function of time for a trap opening time $t_0=1$\,ms, $N=1 \times 10^4$ atoms, and for frequencies ranging from $\omega=2\pi \times 25$\,Hz to $\omega=2\pi \times 200$\,Hz.  By letting the cloud expand in the ring, we bring it close to the Gaussian regime of the quasi-one-dimensional system \cite{petrov}. After a period of 1\,s, ${\cal C} (t)$ is appreciable only if 
$\omega \alt 2\pi \times 100$\,Hz. Our result seems to be in contradiction with Ref.\,\cite{fallon} where long coherence times were obtained only for guide frequencies of $\sim 2\pi \times 10$\,Hz. By allowing a waiting time before splitting, the effective frequency at the time of splitting is low because it is renormalized by the expansion parameters.  

In Fig.\,\ref{coherence}(B), we show how the coherence ${\cal C}(t)$ is influenced by the number of atoms for $\omega=2\pi \times 100$\,Hz, with values of $N$ between $1 \times 10^3$ and $1 \times 10^6$ atoms. At t=1\,s, ${\cal C}(t)$ increases from 0.16 to 0.63. Working with larger $N$ is thus more favorable; however, current cloud sizes in ring guides are only in the range of $N=1 \times 10^4-1 \times 10^5$ \cite{pandey}.

In Fig.\,\ref{coherencew100}, we show how the trap opening time $t_0$ and the delay time before splitting $t_s$ impact the coherence. The optimization of coherence is achieved by opening the trap as fast as possible then allowing a waiting time before splitting. When $\omega=2\pi \times 100$\,Hz, 
we can improve the coherence time (defined as ${\cal C}(t)$ dropping to $1/e$) from 0.36\,s for $t_0$=0.01\,s and $t_s$=0.1\,s to 2\,s for $t_0$=0.001\,s and $t_s$=0.3\,s. However, $t_s$ should not be too long as the condensate can expand significantly in the ring and the beam splitting becomes less effective. Hence for very long $t_s$, the gain in ${\cal C} (t)$ must be balanced by the loss in signal since only a reduced fraction of the atoms will be accelerated in the splitting.  In order to reduce the signal loss due to the curvature of the ring guide, rings with a large radius are thus better. 

{\parindent=0pt
\begin{figure*} 
\begin{centering}
   \includegraphics[width=0.55\textwidth]{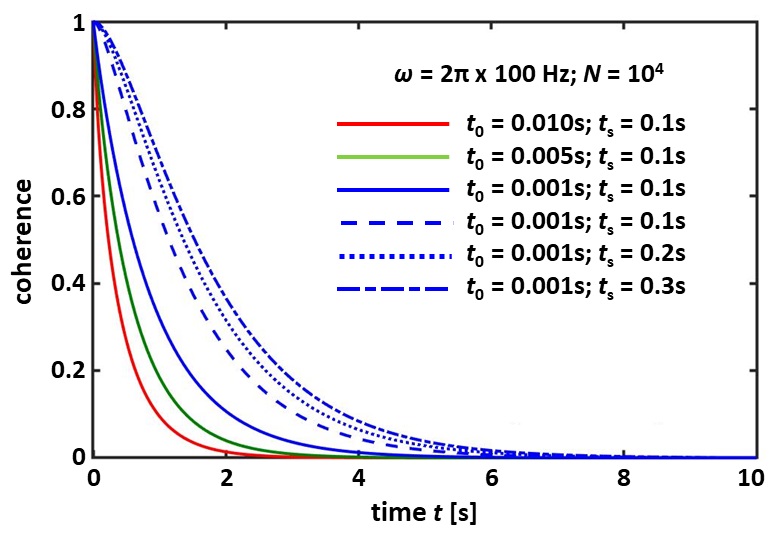} 
	\parbox{0.9\textwidth}{\caption{\label{coherencew100} \baselineskip=0.67\baselineskip Dependence of the coherence on the trap opening time~$t_0$ and the splitting times~$t_s$. Coherence for the best opening time~$t_0=\rm0.001\,s$ is further improved by allowing a waiting time~$t_s$ before splitting. In all cases, the trap frequency is~$\omega=\rm2\pi\times100\,Hz$ and the number of atoms is~$N=10^4$.}} 
\end{centering}
\end{figure*}
}

In this section we have studied the hindering influence of phase diffusion in a BEC confined in a ring trap. We show that coherence is lost rapidly if the guide frequencies are high. We use a dynamical Thomas-Fermi theory to show that coherence can be retained at moderate guide frequencies by letting the BEC expand in the ring before splitting. We show that a coherence time of up to 1\,s can be obtained in a guide with a transverse frequency of $2\pi \times 100$\, Hz by adjusting the trap opening time and the splitting time. Although our analysis was restricted to the Thomas-Fermi limit and the inclusion of the kinetic term would yield more accurate results, it was shown in Ref.\,\cite{jamison} that this correction is small in the high density regime, $N \agt 1 \times 10^4$. The role of interactions on coherence has also been illustrated experimentally in a recent study with a moving linear optical trap \cite{krzyzanowska}. For $^{87}$Rb in which interaction effects are important, the coherence time was 160\,ms, while this was increased to 1\,s by using the weakly interacting $^{39}$K.

\section{Optimized configurations}
\label{configurations}

{\parindent=0pt
\begin{figure*} 
\begin{centering}
   \includegraphics[width=0.9\textwidth]{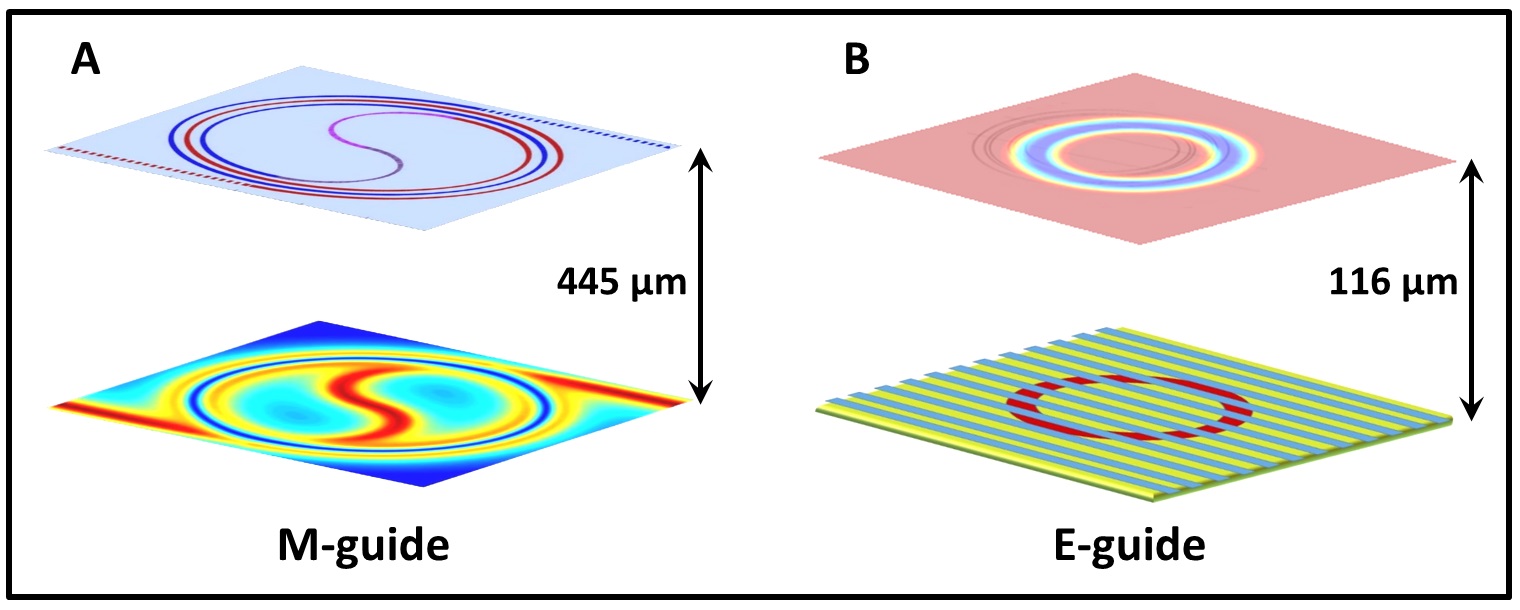} 
	\parbox{0.9\textwidth}{\caption{\label{guides-fig} \baselineskip=0.67\baselineskip Optimum configurations for the M-guide and the E-guide. (A)~Archimedean spiral (M-guide). The upper panel shows the atom chip circuits (see Fig.\,\ref{archi1-fig}), while the lower panel displays the resulting potential guide (blue circle). The atom chip is in the conventional downward orientation. Circuit characteristics: current~$I=\rm1\,A$; TOP bias field,~$\rm1.8\,G$ (the TOP coils are not shown). Guide characteristics: distance below the chip,~$\rm445\,\mu m$; trap depth,~$\rm10\,\mu K$; frequencies,~$\rm2\pi\times120\,Hz$ (radial) and~$\rm2\pi\times175\,Hz$ (axial). (B)~Sketch of the E-Guide. Here we use an inverted chip orientation to minimize the guide frequencies. The lower panel shows the atom chip (see Fig.\,\ref{mirror-fig}), while the upper panel displays the resulting potential guide. Chip characteristics: capacitor voltage,~$\rm80\,V$; magnetic field at the surface of the mirror,~$\rm124\,G$; magnetic domain periodicity, $\rm100\,\mu m$. Guide characteristics: distance above the chip,~$\rm116\,\mu m$; trap depth,~$\rm13\,\mu K$; trap bottom,~$\rm0.085\,G$; frequencies~$\rm2\pi\times60\,Hz$ (radial) and~$\rm2\pi\times196\,Hz$ (axial).}}
\end{centering}
\end{figure*}
} 

In the previous sections, we studied the influence of roughness and phase diffusion on the coherence in a ring guide. Since current fabrication methods can yield a roughness amplitude in the range $\delta u$=0.01-0.1 \textmu m, potential fluctuations at 100\,\textmu m from the chip are in the range of 400\,nK-4\,\textmu K, and atoms can travel along the ring without coherence loss for at least 1\,s if their velocity is larger than 4\,$\hbar k$. It should be noted that a velocity of $2\,\hbar k$ is already sufficient if $\delta u$=0.01 \textmu m. Phase diffusion effects constrain the average guide frequency to $ \sim 2\pi \times 100$\,Hz. Furthermore, the ring radius should be large enough in order to avoid the loss of signal or the collision of the atoms on the guide wall during splitting.   These considerations lead us to choose a guide radius of 5\,mm. Larger guides are better in principle  because they have larger areas. However, since the two wavepackets must travel further from each other before recombination, decoherence due to the environment could become another hindering effect. Finally, the effects of gravity and centrifugal force must be included.  

The total energy of the guide is written as

\begin{equation}
U=V_{\text{guide}}\pm mg(z-z_0)+m \frac {v^2}{r_0}(r-r_0)+V_{\text{ext}},
\label{guiden-eq}
\end{equation}

\noindent where $V_{\text{guide}}$ is the guide potential excluding the gravity and centrifugal potentials, $v$ is the velocity of the atoms, and $r_0$ the radius of the ring. $V_{\text{ext}}$ is an external potential, for instance the TOP potential applied to the M-guide. For the M-guide, $V_{\text{guide}}=\mu_B |B|$ and for the E-guide, $V_{\text{guide}}=V_{\text{capacitor}}+B_0 \exp(-2\pi z/a)$, where $B_0$ is the magnetic field at the chip surface, $z$ is the distance from that surface, and $a$=100\,\textmu m is the period of the magnetic mirror. Choosing a large radius for the M-guide was particularly important because the guide has to be far enough from the chip in order to significantly reduce the frequencies and the bumps due to the current leads. This is because in the Archimedean circuit, the guide position is related to its radius. Furthermore, we needed to add the TOP potential in order to get rid of the potential zeros, and we set the TOP field to 1.8\,G.  Finally, while for the M-guide we adopted the conventional orientation for which the guide is below the chip, the guide is instead above the chip for the E-guide in order to minimize the guide frequencies. 

\begin{table*}
\makegapedcells
\begin{center}
\begin{tabular}{ | c | c | c | c | c | c | c |}
\hline
 \multicolumn{7}{|c|}{M-guide}\\ 
\hline
~~Height~~& ~~Depth~~ & TOP bias field & Axial freq. & Radial freq. & Coherence time & Max. velocity\\ 
\hline
445\,\textmu m & 10\,\textmu K & 1.8\,G & 2$\pi \times$175\,Hz & 2$\pi \times$120\,Hz & 0.51\,s & 32\,$\hbar k$\\ 
\hline
 \multicolumn{7}{|c|}{E-guide}\\
\hline
~~Height~~& ~~Depth~~ & Trap bottom & Axial freq. & Radial freq. & Coherence time& Max. velocity\\ 
\hline
116\,\textmu m & 13\,\textmu K & 0.085\,G & 2$\pi \times$196\,Hz & 2$\pi \times$60\,Hz & 0.87\,s & 30\,$\hbar k$\\
\hline
\end{tabular}
\parbox{0.9\textwidth}{\caption{\baselineskip=0.67\baselineskip Optimum parameters for the M-guide with $r_0$=5 mm. $I$=1\,A, the trap opening time was $t_0$=1\,ms, and there was no waiting time before splitting. Optimum parameters for the E-guide with $r_0$=5 mm. Capacitor voltage: 80\,V, magnetic field at the surface of the mirror: 124\,G, mirror periodicity: 100\,\textmu m. The trap opening time was $t_0$=1\,ms, and there was no waiting time before splitting.}}
\label{guides-tbl}
\end{center}
\end{table*}

When searching for the best guide configuration, it is necessary to address the following problem related to the guide stability and coherence. If the wavepacket energy is high enough, it will propagate in the ring guide without loss within the coherence time allowed by phase diffusion. Performing many turns increases the effective area of the interferometer and yields high sensitivity. On the other hand, in high frequency guides where high energy wavepackets can be sustained, the interactions between atoms are important and lead to phase diffusion.

Our optimized configurations are shown in Fig.\,\ref{guides-fig} and the corresponding parameters are shown in Table\,\ref{guides-tbl} for the M-guide and for the E-guide. 

{\parindent=0pt
\begin{figure*} 
\begin{centering}
   \includegraphics[width=0.9\textwidth]{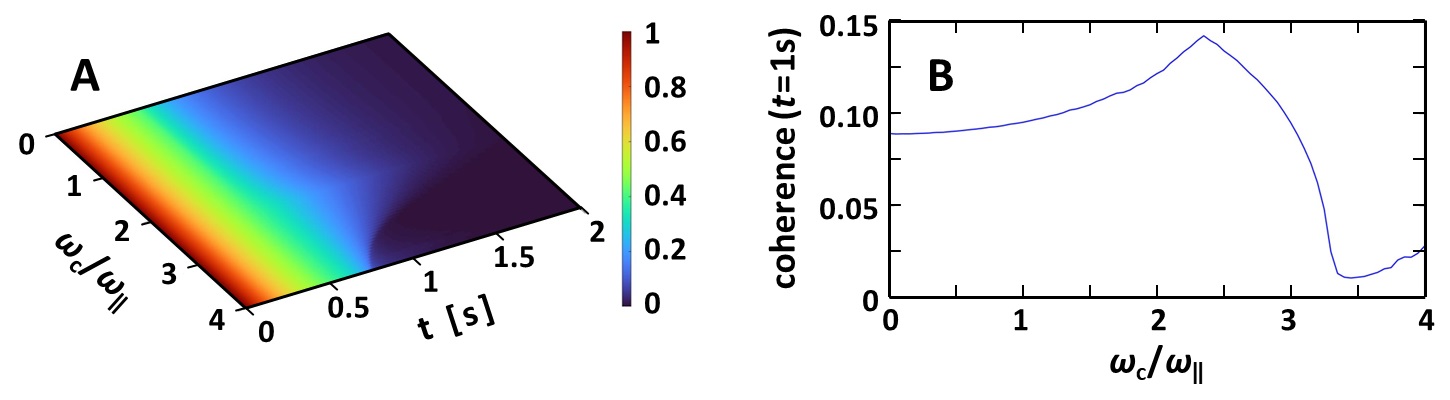} 
	\parbox{0.9\textwidth}{\caption{\label{collimation-fig} \baselineskip=0.67\baselineskip Influence of focusing on coherence: (A)~Coherence as a function of time and the ratio~$\omega_c/\omega_\parallel$ of the focusing frequency to the parallel frequency of the optical dipole trap for a focusing time of~$t_c=\rm1\,ms$. (B)~Cut through~(A) at~$t=\rm1\,s$. The effective coherence is enhanced for focusing frequencies below~$2.5\,\omega_\parallel$ and has a maximum near~$\omega_c/\omega_\parallel=2.5$.}} 
\end{centering}
\end{figure*}
}

\section{Projected sensitivities}
\label{sensitivities}

The interferometer sensitivity $\sigma$ is obtained from the observed signal which is the atom population,

\begin{equation}
 P=\frac{1}{2}\left[1-\nu \cos(\Phi + \phi_0)\right],
\end{equation}

\noindent where $\nu$ is the interferometer visibility, $\Phi=2m{\tilde A}\Omega/\hbar$ is the Sagnac phase, $\Omega$ is the angular speed, and $\phi_0$ is a reference phase. $\sigma$ reads \cite{garrido-alzar},

\begin{equation}
  \sigma=\frac{\hbar}{m\nu{\tilde A}\sqrt{N}},
\label{sensitivity-eq}
\end{equation}

\noindent where ${\tilde A}$ is the effective area, which is the actual area multiplied by the number of revolutions, $\nu$ is the the interferometer visibility, and the factor $\sqrt{N}$ is the signal-to-noise ratio in the quantum limit.

A realistic estimate of sensitivities is obtained by taking into account the effects of roughness and phase diffusion on the visibility $\nu$. We find that if the roughness amplitude $\delta u \lesssim$ 0.01\,\textmu m, no reduction of visibility occurs. Assuming this value for the roughness, we now analyze the influence of phase diffusion. The BEC is loaded into the ring guide from an optical dipole trap whose frequencies $(\omega_{\parallel}, \omega_{\perp})$ are chosen to allow adiabatic loading. For instance, for the E-guide, we can choose ($\omega_{\parallel}, \omega_{\perp})=(2\pi \times 25\,\text{Hz}, 2\pi \times 130$\,Hz) so that the transverse frequency of the dipole trap is midway between the axial ($2\pi \times 196$\,Hz) and radial ($2 \pi \times 60$\,Hz) frequencies of the guide. After release from the trap, the BEC is allowed to expand for 1\,ms and then it is split. After an evolution of 1\,s in the ring trap, each half of the BEC will have expanded to a length of 1800\,\textmu m. This may exceed the field-of-view of the detection optics and camera, leading to atom loss and signal reduction.

This atom loss can be avoided by applying a focusing pulse \cite{abend_2}, i.e. a brief re-trapping of the atoms halfway during their revolutions. The disadvantage of applying such a focusing pulse is that it leads to an increase of the average trap frequency felt by the atoms and thereby to an increase of the phase diffusion rate. In Fig.\,\ref{collimation-fig}, the coherence is shown as a function of time and the focusing frequency, with the focusing time set to $t_c$=1\,ms. The coherence is adjusted by including a factor related to the number of atoms; letting $\ell_c$ be the width of the camera field-of-view (we assume a typical value of $\sim$ 700 \,\textmu m), we multiply the coherence by $\ell_c/\text{max}(\ell_c, \ell(t))$, where $\ell(t)$ is the length of the cloud at time $t$. Fig.\,\ref{collimation-fig} shows that the effective coherence is enhanced for focusing frequencies below $2.5 \,\omega_{\parallel}$.

\begin{table*}
\makegapedcells
\begin{center}
\begin{tabular}{ | c | c | c | c  | c | c | c |} 
\hline
& ~Stanford~ & Hannover & ~~ Paris~~  & M-guide & E-guide & E-guide-2\\ 
\hline
Flux($\text{s}^{-1})$ & $ 10^{10}$ & $ 10^9 $ &$2\,10^7$ & $10^4$ & $ 10^4$ & $ 10^{4}$\\
\hline
Effective area ($\text{mm}^2$) & $24$ & $41$ & $1100$ & $471$ & $471$ & $471$\\
\hline
Sensitivity($\text{nrad}\,\text{s}^{-1}\,\text{Hz}^{-1/2}$) & $ 0.6 $ & $120$ & $100$ & $241$ & $78.7$ & $45.4$\\
\hline
Interferometer length (m)& 0.97 & 0.14 & 0.58 & 4.45 10$^{-4}$& 1.16 $10^{-4}$ &1.16 $10^{-4}$ \\
\hline
Sensitivity $\times$ length ($\text{nrad}\,\text{s}^{-1}\,\text{Hz}^{-1/2}\,$m) & $ 0.58 $ & $16.4$ & $58$ & $0.12$ & $0.008$ & $0.005$\\
\hline
\end{tabular}
\parbox{0.9\textwidth} {{\caption{\baselineskip=0.67\baselineskip Projected sensitivities for the M-guide, the E-guide, and the E-Guide with focusing (E-Guide-2) are compared to the state-of-the-art free-space sensitivity data from Stanford \cite{gustavson}, Hannover \cite{berg}, and Paris (Syrte) \cite{dutta}. The effective area for the M-guide and E-guide corresponds to 6 revolutions. Although the initial conditions are identical for the M-guide and the E-guide, the difference in their sensitivities stems from their different trap frequencies and the application of the focusing pulse. In the last line, we display the product of the sensitivity and the length of the interferometer, a measurement of sensitivity that includes the compactness of the interferometer. Based on this figure of merit, the E-Guide is better than the free-space interferometers by 2-4 orders of magnitude.}}
\label{sensitivity-tbl}}
\end{center}
\end{table*}

In Table\,\ref{sensitivity-tbl}, we show projected sensitivities for the M-guide and the E-guide with and without focusing, which we compare to data from Paris (Syrte) and Stanford (taken from Ref.\,\cite{barrett}). The effective area in the M-guide and E-guide is obtained for 6 revolutions. The projected sensitivities are estimated by taking into account coherence losses induced by roughness and phase diffusion. Additional losses due to other sources of technical noise are added by taking the geometric average of these losses in the Paris \cite{gauguet} and Stanford \cite{gustavson} experiments. The losses in these experiments were estimated by comparing the experimental value of the sensitivity in each experiment with the corresponding ideal sensitivity obtained from Eq.\,\ref{sensitivity-eq}. This geometric average added a factor of 7 to the sensitivity. It can be seen in Table\,\ref{sensitivity-tbl} that, despite having a flux of three and six orders of magnitude less than the Paris and Stanford experiments respectively, the projected sensitivities are comparable to the Paris sensitivity and only two orders of magnitude smaller than the Stanford sensitivity. The last two columns of Table\,\ref{sensitivity-tbl} shows that the focusing yields only a modest gain for the sensitivity.

\section{ Refinement: quantum revivals and larger rings}
\label{revivals}

Despite the BEC having a typical trap lifetime of about 10\,s, we have seen that phase diffusion restricts the interferometer time to a duration of less than 1\,s. This time limits the interferometer sensitivity. In order to improve the sensitivity, we can take advantage of the ability of quantum mechanical systems to show revivals \cite{robinett}, i.e, after a period, $\tau_r$, an apparently lost coherence is restored. Revivals were predicted in BEC's \cite{wright} and observed experimentally \cite{greiner}. There is also a revival time related to phase diffusion \cite{javanainen},  $\tau_r=2\pi \sqrt{N}/R$, where $N$ is the number of atoms and $R$ is the phase diffusion rate. It is thus possible to adjust the trap opening time and the splitting time to set the revival time to $\tau_r=10$\,s for the 5\,mm radius interferometer. Hence we can consider allowing the atom clouds to perform 52 revolutions (much longer than the 6 revolutions considered without revivals) when the momentum imparted by the beam splitter is 30\,$\hbar k$. Applying the focusing scheme above will now yield a sensitivity of 0.63 nrad\,s$^{-1}$. If the same factor 7 as above is included for operational losses, the interferometer will yield a sensitivity of 4.4 nrad\,s$^{-1}$.   

Finally, we estimate the best sensitivity attainable with currently available technology. The flux in our study was restricted to $1 \times 10^4$ atoms, which has been achieved in our group. But some groups have reported BEC's of up to $3 \times 10^5$ atoms \cite{pandey}. Furthermore, since a wavepacket separation on a half-meter scale has been realized \cite{kovachy}, a ring radius of 50\,mm is achievable. Due to the large radius, the effects of the centrifugal force are reduced and the guides can sustain higher momenta.  Taking for instance 60\,$\hbar k$ and a BEC of $3 \times 10^5$ atoms, and assuming the same operational losses, this guide configuration will yield a sensitivity of 5 prad\,$\text{s}^{-1}$ after 10\,s. This is better than the sensitivity, 10 prad\,$\text{s}^{-1}$ after 120\,s, obtained from a large-scale ring gyroscope the size of a five-story building \cite{gebauer}. Furthermore, the calculation of phase diffusion rate used the time-dependent Thomas-Fermi model which neglects the transverse kinetic energy and may allow the transverse size of the BEC to shrink below the waveguide ground-state size. An improved calculation that takes the kinetic energy into account~\cite{Japha2019} would reduce the rate of phase diffusion and improve the sensitivity.
 
We may alternatively consider using thermal atoms instead of a BEC. Proof-of-principle of such a high-sensitivity interferometer was demonstrated in \cite{yoni}. This interferometer is operated in a continuous way with a thermal beam formed by a 2D MOT, utilizing tunneling barriers as the beam splitters. In this scenario we may fully benefit from the long vacuum-limited lifetime of about 10\,s  and of the higher speed of cold atoms in the ring trap. This interferometer configuration, with an assumed atomic flux of $\text{10}^{\text{7}}\,\text{s}^{\text{-1}}$, is predicted to yield a sensitivity of $\text{1}\, \text{prad}\,\text{s}^{\text{-1}}\,\text{Hz}^{\text{-1/2}}$ for 1000 turns. Again, considering the same figure of merit, the improvement over the state-of-the-art for a thermal beam is expected to be 4-6 orders of magnitude.

\section{ Conclusion}

Guided atom interferometers are expected to solve some of the limitations in sensitivity and compactness that hinder the development of free-space atom interferometers. Different platforms for guided Sagnac atomic interferometers are currently being developed around the world. These guided Sagnac interferometers are based on a variety of magnetic, optical, or magneto-optical potentials. Despite these efforts however, a coherent signal has been obtained only in the special case of a moving linear trap \cite{wu,tonyushkin,krzyzanowska}.  In this study we analyze two intrinsic factors that distinguish guided interferometry from free-space interferometry, namely guide roughness and phase diffusion due to effects of atom confinement leading to larger atom-atom interactions. Our analysis shows that despite the current hindrances there is no fundamental impediment to the realization of high-sensitivity guided atom gyroscopes.

Improved fabrication methods and time-averaged potentials can yield extremely smooth guides. However, such time-dependent potentials can themselves induce coherence loss. Guides generated by static fields thus appear to be attractive alternatives thanks to their flexibility in creating guides of different radii.  

Using the time-dependent Thomas-Fermi approximation, we show how the presence of atom-atom interactions radically differentiate guided interferometry from free-space interferometry. In free space, after release from the trap, the atoms can expand in the three spatial directions so that effects of atom-atom interactions quickly become negligible. But in a guide, the atoms are free to expand only in the azimuthal direction and atom-atom interactions therefore play a critical role. They significantly reduce the coherence time in
the ring trap. In this study, we devised a method to increase the coherence time long enough to allow for interferometry in a ring guide.
    
Finally, owing to recent progress in achieving high-momentum beam splitting and in constructing smooth guides, high-sensitivity guided interferometry can be implemented within the coherence time allowed by phase diffusion. Despite lower fluxes in guided Sagnac interferometers, sensitivities comparable to those obtained with free-space interferometers can be obtained provided that atom velocities in the guide are $\sim 30 \hbar k$. Higher sensitivities can be achieved by using the possibility of quantum revivals and large-scale separation. Assuming the guide to have a radius of 50\,mm, and the momentum imparted to the atoms $\sim 60 \hbar k$, both values achievable with current technologies,  sensitivities higher than the state-of-the-art sensitivity of giant ring laser gyroscopes can be achieved. 

\begin{acknowledgments}
We would like to thank Peter Hannaford (Swinburne University of Technology) for helpful discussions regarding magnetic mirrors and for sending us realistic parameters. This paper is based upon research supported in part by the Israel Science Foundation (ISF) grants No. 856/18 and No. 1314/19. We also acknowledge support from the Israeli Ministry of Immigrant Absorption (MK).   
\end{acknowledgments}

\end{document}